\documentclass[twocolumn,showpacs,raggedbottom,raggedfooter,amsmath,amssymb,pra,citeautoscript,a4paper]{revtex4}
\usepackage{amsmath}
\usepackage{amsfonts}
\usepackage{amssymb}

\newcommand{\tr}{\mathrm{tr}}

\newcommand{\1}{{\rm 1\hspace{-0.9mm}l}}

\newtheorem{theorem}{Theorem}
\newtheorem{lemma}{Lemma}

\newcommand{\halmos}{\vspace{3mm}\hfill $\blacksquare$}
\newcommand{\proof}{\noindent {\it Proof.\ }}
\newcommand{\proofof}[1]{\noindent {\it Proof of #1.}}

\newcommand{\ket}[1]{| #1 \rangle}

\newcommand{\ketbra}[2]{| #1 \rangle \langle #2 |}

\newcommand{\Cplx}{\mathbb{C}}

\usepackage{graphicx}

\begin{document}

\title{Bound on trace distance based on superfidelity}

\author{Zbigniew Pucha{\l}a}
\email{z.puchala@iitis.gliwice.pl}
\affiliation{Institute of Theoretical and Applied Informatics, Polish Academy of 
Sciences, Ba{\l}tycka 5, 44-100 Gliwice, Poland}

\author{Jaros{\l}aw Adam Miszczak}
\affiliation{Institute of Theoretical and Applied Informatics, Polish Academy
of Sciences, Ba{\l}tycka 5, 44-100 Gliwice, Poland}
\affiliation{Faculty of Informatics, Masaryk University, Botanick\'{a} 68a,
602\,00 Brno, Czech Republic}

\begin{abstract}
We provide a bound for the trace distance between two quantum states. The 
lower bound is based on the superfidelity, which provides the upper bound on 
quantum fidelity. One of the advantages of the presented bound is that it can 
be estimated using a simple measurement procedure. We also compare this bound 
with the one provided in terms of fidelity. 
\end{abstract}
\date{11 February 2009}
% quantum information, quantum mechanics, matrix theory
\maketitle

%%%%%%%%%%%%%%%%%%%%%%%%%%%%%%%%%%%%%%%%%%%%%%%%%%%%%%%%%%%%%%%%%%%%%%%%%%%%%%%%
\section{Introduction}
%%%%%%%%%%%%%%%%%%%%%%%%%%%%%%%%%%%%%%%%%%%%%%%%%%%%%%%%%%%%%%%%%%%%%%%%%%%%%%%%
The trace distance is one of the most natural distance measures used in 
quantum-information theory \cite{BZ06,NC00}. It is one of the main tools used in 
distinguishability theory and it is connected to the average success 
probability when distinguishing two states by a measurement \cite{ziman,En96,FG99}. 
It is also related to quantum fidelity, which provides the measure of 
similarity of two quantum states \cite{BZ06,jozsa94,MMPZ08}. 

Both quantities are particularly important in quantum cryptography, since the 
security of quantum protocols relies on an ability to measure the distance 
between two quantum states \cite{FG99}. The trace distance is also related to 
other properties of quantum states like the von Neumann entropy and relative 
entropy~\cite{BZ06}.

The main aim of this work is to provide a lower bound for the trace distance using 
measurable quantities. We use nonlinear functions of the form 
$\tr\rho_i\rho_j, i,j=1,2$, where $\rho_i$ and $\rho_j$ are density matrices. 
For such forms there exist feasible schemes to measure them in an experiment without 
resorting to state tomography \cite{bovino}. 
We give a lower bound based on the superfidelity introduced recently in 
\cite{subsuper}. Proof of this bound gives an answer to the conjecture stated by 
Mendonca {\it et al.} in \cite{mendonca}.

%%%%%%%%%%%%%%%%%%%%%%%%%%%%%%%%%%%%%%%%%%%%%%%%%%%%%%%%%%%%%%%%%%%%%%%%%%%%%%%%
\section{Bounds on trace distance}
%%%%%%%%%%%%%%%%%%%%%%%%%%%%%%%%%%%%%%%%%%%%%%%%%%%%%%%%%%%%%%%%%%%%%%%%%%%%%%%%
Let us denote by $\Omega_N$ the space of density matrices acting on 
$N$-dimensional Hilbert space $\Cplx^N$. For two density matrices $\rho_1, \rho_2\in\Omega_N$
the trace distance is defined as
\begin{equation}
D_{\tr}(\rho_1,\rho_2)=\frac{1}{2}\tr|\rho_1-\rho_2|.
\end{equation}
In the particular case of pure states we can use Bloch vectors
$\rho_1=\vec{r}\cdot\vec{\sigma}$ and $\rho_2=\vec{s}\cdot\vec{\sigma}$.
One can see that the trace distance between such states is equal to half of the 
Euclidean distance between the respective Bloch vectors
\begin{equation}
D_\tr(\rho_1,\rho_2)=\frac{|\vec{r}-\vec{s}|}{2}.
\end{equation}

The trace distance can be bounded with the use of the fidelity~\cite{FG99,BZ06}
\begin{equation}\label{eqn:fid-dtr-ineq}
1-\sqrt{F(\rho_1,\rho_2)}\leq D_{\tr}(\rho_1,\rho_2)\leq\sqrt{1-F(\rho_1,\rho_2)},
\end{equation}
where the fidelity is defined as 
\begin{equation}\label{eqn:fid}
F(\rho_1,\rho_2)=\left[\tr |\sqrt{\rho_1}\sqrt{\rho_2} | \right]^2.
\end{equation}
The inequality (\ref{eqn:fid-dtr-ineq}) shows that $F$ and $D_\tr$ are closely
related indicators of distinguishability.

The main result of this work is a lower bound for the trace distance, which 
we prove in the next section, 
\begin{equation}\label{eqn:Main-ineq}
1-G(\rho_1,\rho_2)\leq D_{\tr}(\rho_1,\rho_2),
\end{equation}
where $G$ is called the \emph{superfidelity}, and was introduced in \cite{subsuper}.
For $\rho_1,\rho_2\in\Omega_N$ it is defined as
\begin{equation}
G(\rho_1,\rho_2) = \tr\rho_1\rho_2 + \sqrt{1-\tr \rho_1^2}\sqrt{1-\tr \rho_2^2}.
\end{equation}
From the matrix analytic perspective the inequality (\ref{eqn:Main-ineq})
relates the trace norm on the space $\Omega_N$ to the Hilbert-Schmidt scalar product 
on $\Omega_N$. %Supplementary connection was provided in \cite{QCB} in a context 
%of quantum Chernoff bound.

For the sake of consistency we provide basic information about 
the superfidelity~\cite{subsuper}. The most interesting feature of superfidelity
is that it provides an upper bound for quantum fidelity~\cite{subsuper}
\begin{equation}
F(\rho_1,\rho_2)\leq G(\rho_1,\rho_2).
\end{equation}

The superfidelity also has properties which make it useful for quantifying the 
distance between quantum states. In particular we have: (1) Bounds: $0 \le G(\rho_1,\rho_2) \le 1$.
(2) Symmetry: $G(\rho_1,\rho_2)= G(\rho_2,\rho_1)$.
(3) Unitary invariance: for any unitary operator $U$, we have $G(\rho_1,\rho_2)=G(U\rho_1U^{\dagger},U\rho_2U^{\dagger})$.
(4) Concavity:   
$G(\rho_1,\alpha \rho_2 + (1-\alpha)\rho_3) \geq \alpha G(\rho_1,\rho_2) + (1-\alpha) G(\rho_1,\rho_3)$ 
for any $\rho_1,\rho_2,\rho_3\in\Omega_N$ and $\alpha \in [0,1]$.
(5) Supermultiplicativity: for 
$\rho_1,\rho_2,\rho_3,\rho_4 \in \Omega_N$ we have
\begin{equation}
G(\rho_1 \otimes \rho_2, \rho_3 \otimes \rho_4) \geq G(\rho_1,\rho_3) G(\rho_2,\rho_4).
\end{equation}

Note that the superfidelity shares properties 1-4 with the fidelity. However, in 
contrast to the fidelity, superfidelity is not multiplicative, but 
supermultiplicative.

In \cite{mendonca} the authors showed that $G$ is \emph{jointly} concave in 
its two arguments, 
\begin{eqnarray}
G\big(\alpha\rho_1+(1-\alpha)\rho_2\!\!\!&,\alpha\rho'_1+(1-\alpha)\rho'_2\big)\\\nonumber 
\geq \alpha G(\rho_1,\rho'_1)+&\!\!\!(1-\alpha)G(\rho_2,\rho'_2), 
\end{eqnarray}
for $\alpha \in [0,1]$. Note that the property of joint concavity is obeyed
by the square root of fidelity $\sqrt{F}(\rho_1,\rho_2)=\tr|\sqrt{\rho_2}\sqrt{\rho_2}|$,
but not by the fidelity (\ref{eqn:fid}).

Fidelity can be used to define the metric $D_B(\rho_1,\rho_2)$ on the space 
$\Omega_N$ as
\begin{equation}
  D_B(\rho_1,\rho_2) = \sqrt{ 2 - 2 \sqrt{F(\rho_1,\rho_2)}} .
 \label{eqn:bures}
\end{equation}
Unfortunately the analog of the Bures distance defined using the superfidelity
\begin{equation}
  D_B'(\rho_1,\rho_2) = \sqrt{ 2 - 2 \sqrt{G(\rho_1,\rho_2)}}
 \label{eqn:not-metric-super}
\end{equation}
is not a metric \cite{mendonca}, but the quantity 
\begin{equation}
  D_G(\rho_1,\rho_2) = \sqrt{ 2 - 2 G(\rho_1,\rho_2)}
 \label{eqn:metric-super}
\end{equation}
provides the metric on the space $\Omega_N$.

Finally one should note that the superfidelity is particularly convenient to use
as the practical measure of similarity between quantum states. One of the main 
advantages of superfidelity is that it is possible to design feasible schemes 
to measure it in an experiment \cite{subsuper}. Also, from the computational 
point of view, calculation of the superfidelity is significantly less 
resource-consuming \cite{mendonca}.

%%%%%%%%%%%%%%%%%%%%%%%%%%%%%%%%%%%%%%%%%%%%%%%%%%%%%%%%%%%%%%%%%%%%%%%%%%%%%%%%
\section{Main result}
%%%%%%%%%%%%%%%%%%%%%%%%%%%%%%%%%%%%%%%%%%%%%%%%%%%%%%%%%%%%%%%%%%%%%%%%%%%%%%%%
The properties of superfidelity listed above suggest that it would be convenient 
to use it instead of fidelity to draw conclusions about the distinguishability
of quantum states. This section show how this can be done by relating the 
superfidelity and trace distance.

First we can observe that from the inequality
\begin{equation}
F(\rho_1,\rho_2)\leq G(\rho_1,\rho_2),
\end{equation}
and since we have (\ref{eqn:fid-dtr-ineq}) we get that
\begin{equation}
1-\sqrt{G(\rho_1,\rho_2)}\leq D_{\tr}(\rho_1,\rho_2).
\end{equation}
Our main aim is to prove the following inequality, which provides tighter bound.
\begin{theorem}\label{conjecture-general}
For any $\rho_1, \rho_2\in\Omega_N$ we have
\begin{equation}
1-G(\rho_1,\rho_2)\leq D_{\tr}(\rho_1,\rho_2),
\end{equation}
or equivalently 
\begin{eqnarray}\label{eqn:conjecture-general}
\frac{1}{2} \tr |\rho_1 - \rho_2| + \tr \rho_1\rho_2 + 
\sqrt{1-\tr \rho_1^2}\sqrt{1-\tr \rho_2^2} \geq 1.
\end{eqnarray}
\end{theorem}
This inequality was first stated as a conjecture in \cite{mendonca}, 
where it was verified numerically for small dimensions. Clearly it 
is motivated by the lower bound for trace distance provided by the inequality 
(\ref{eqn:fid-dtr-ineq}). 

To prove the Theorem \ref{conjecture-general} we need the following lemma.
\begin{lemma}
For any $\rho_1, \rho_2\in\Omega_N$ let $P_{+}$ and $P_{-}$ be the 
projectors onto $(\rho_1-\rho_2)^{+}$ and $(\rho_1-\rho_2)^{-}$ respectively. 
We have the following inequalities
\begin{eqnarray}
  \tr P_{+} (\1 - \rho_1) \rho_1 &\geq& \tr P_{+} (\1 - \rho_1) \rho_2, \label{ineq:1}\\
  \tr P_{-} \rho_1 (\1 - \rho_1) &\geq& \tr P_{-} \rho_1 (\1 - \rho_2), \label{ineq:2}\\
  \tr P_{+} (\1 - \rho_2) \rho_2 &\geq& \tr P_{+} (\1 - \rho_1) \rho_2, \label{ineq:3}\\
  \tr P_{-} \rho_2 (\1 - \rho_2) &\geq& \tr P_{-} \rho_1 (\1 - \rho_2). \label{ineq:4}
\end{eqnarray}
\end{lemma}
\proof
Because of the similarity we will show only inequality (\ref{ineq:1}). It is 
easy to prove inequalities (\ref{ineq:2}), (\ref{ineq:3}) and (\ref{ineq:4})
in a similar manner. 

We subtract the right- from the left-side of (\ref{ineq:1}) to get
\begin{eqnarray}
&& \!\!\!\! \tr P_{+}  (\1 - \rho_1) \rho_1- \tr P_{+} (\1 - \rho_1) \rho_2 \\\nonumber
&&= \tr P_{+} (\1 - \rho_1) (\rho_1-\rho_2) \\\nonumber
&&=
\tr P_{+}(\1 - \rho_1) (\rho_1-\rho_2) P_{+} \geq  0
\end{eqnarray}
because $\1 -\rho_1$ is positive semidefinite.
\halmos

\proofof{Theorem \ref{conjecture-general}}
Adding the inequalities (\ref{ineq:1}) and (\ref{ineq:2}) we get
\begin{equation}\label{eqn:l1}
  \tr \rho_1 (\1 - \rho_1)\geq \tr P_{+} (\1 - \rho_1) \rho_2 + \tr P_{-} \rho_1 (\1 - \rho_2).
\end{equation}
Similarly, by adding inequalities (\ref{ineq:3}) and (\ref{ineq:4}) we get
\begin{equation}\label{eqn:l2}
  \tr \rho_2 (\1 - \rho_2)\geq \tr P_{+} (\1 - \rho_1) \rho_2 + \tr P_{-} \rho_1 (\1 - \rho_2).
\end{equation} 
Now we notice that, if two non-negative numbers are greater than the third one, 
then so is the geometric mean of the first two numbers. Using this fact we 
combine (\ref{eqn:l1}) and (\ref{eqn:l2}) to get
\begin{eqnarray}
&\sqrt{\tr \rho_1(\1-\rho_1)}&\!\!\! \sqrt{\tr \rho_2(\1-\rho_2)} \\\nonumber 
&&\geq \tr P_{+} (\1 -\rho_1) \rho_2 + \tr P_{-}\rho_1 (\1 -\rho_2).
\end{eqnarray}
On the other hand we can rewrite the trace distance with the use of projectors $P_+$
and $P_-$:
\begin{eqnarray}\label{eqn:Dtr_Projectors}
&D_{\tr}&\!\!\!\!(\rho_1,\rho_2)= \\\nonumber 
&&= \frac{1}{2} (\tr P_{+}\rho_1 - \tr P_{+}\rho_2 + \tr P_{-}\rho_2 - \tr P_{-}\rho_1) \\\nonumber
&&= \frac{1}{2} (\tr P_{+}\rho_1 + \tr P_{+}\rho_2 + \tr P_{-}\rho_2 + \tr P_{-}\rho_1) \\\nonumber 
&&\;\;\:\:-\; \tr P_{+}\rho_2  - \tr P_{-}\rho_1\\\nonumber
&&= 1 - \tr P_{+}\rho_2 - \tr P_{-}\rho_1 .
\end{eqnarray}
Now finally we can write
\begin{eqnarray}
&\frac{1}{2}&\!\!\! \tr |\rho_1 - \rho_2| + \sqrt{1-\tr \rho_1^2}\sqrt{1-\tr \rho_2^2} \\\nonumber
%&&= 1 - \tr P_{+}\rho_2 - \tr P_{-}\rho_1 + \sqrt{1-\tr \rho_1^2}\sqrt{1-\tr \rho_2^2} \\\nonumber
&&\geq 1 - \tr P_{+}\rho_2 - \tr P_{-}\rho_1 \\\nonumber
&&\;\;\:\:+\; \tr P_{+} (\1 -\rho_1)\rho_2 + \tr P_{-}\rho_1 (\1 -\rho_2)\\\nonumber
%&&= 1 - (\tr P_{+}\rho_1\rho_2+\tr P_{-}\rho_1\rho_2)\\\nonumber
%&&=\frac{1}{2} (\tr (P_{+}+P_{-})\rho_1 + \tr (P_{+}+P_{-}) \rho_2) -\tr \rho_1\rho_2 \\\nonumber
&&= 1 -\tr \rho_1\rho_2,
\end{eqnarray}
which is equivalent to (\ref{eqn:conjecture-general}).
\halmos

%%%%%%%%%%%%%%%%%%%%%%%%%%%%%%%%%%%%%%%%%%%%%%%%%%%%%%%%%%%%%%%%%%%%%%%%%%%%%%%%
\section{Comparison of bounds}
%%%%%%%%%%%%%%%%%%%%%%%%%%%%%%%%%%%%%%%%%%%%%%%%%%%%%%%%%%%%%%%%%%%%%%%%%%%%%%%%
It is natural to consider the relation between the lower bounds in 
(\ref{eqn:fid-dtr-ineq}) and (\ref{eqn:Main-ineq}).
It is clear that bound (\ref{eqn:Main-ineq}) is better than 
(\ref{eqn:fid-dtr-ineq}) whenever $G-\sqrt{F} \leq 0$.

In the one qubit case or if one of the states is pure the bound 
(\ref{eqn:Main-ineq}) is always better than (\ref{eqn:fid-dtr-ineq}). 
This follows from the equality between fidelity and superfidelity in these 
situations~\cite{subsuper}. On the other hand the inequality 
(\ref{eqn:fid-dtr-ineq}) provides a better lower bound if the states $\rho_1$ 
and $\rho_2$ have orthogonal supports. In this case the fidelity between states 
vanishes, but the superfidelity is not necessarily equal to zero~\cite{subsuper}.

To get some feeling about the difference between the bound given by fidelity and 
the present bound we will consider the following families of states.
\begin{enumerate}
 \item The family $\rho_{\alpha}$ is defined as
\begin{equation}\label{eqn:ex-states}
\rho_\alpha=\alpha\ketbra{\psi}{\psi}+(1-\alpha)\1/N,
\end{equation}
where $\ketbra{\psi}{\psi}$ is a pure state and in our case we take 
$\ket{\psi}=\ket{0}\in \Cplx^{N}$. 

\item The family $\sigma_{\beta} \in \Omega_8$ is defined as
\begin{equation}\label{eqn:ex-states-dim4}
\sigma_\beta=\beta\ketbra{GHZ}{GHZ}+(1-\beta)\1/N,
\end{equation}
where $\ket{GHZ}=(1/\sqrt{2})(\ket{000}+\ket{111})$.

\item The family $\tau_{\gamma} \in \Omega_8$ is defined as:
\begin{equation}\label{eqn:ex-states-k3}
\tau_\gamma=\gamma\ketbra{010}{010}+(1-\gamma)\1/N.
\end{equation}
\end{enumerate}

\begin{figure}[ht!]
\includegraphics[scale=0.7]{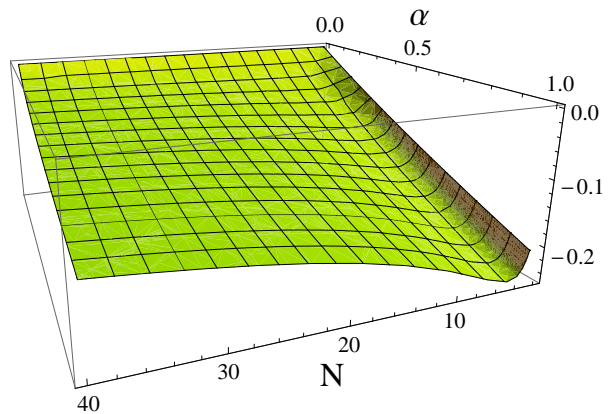}
\caption{The difference $G(\rho_\alpha,\1/N)-\sqrt{F(\rho_\alpha,\1/N)}$ 
calculated for states (\ref{eqn:ex-states}) and maximally mixed states as a 
function of the dimension $N$ and the parameter $\alpha$ 
[see Eq.~(\ref{eqn:ex-states})].}
\label{fig:low-bound-diff}
\end{figure}
\begin{figure}[ht!]
\includegraphics[scale=0.7]{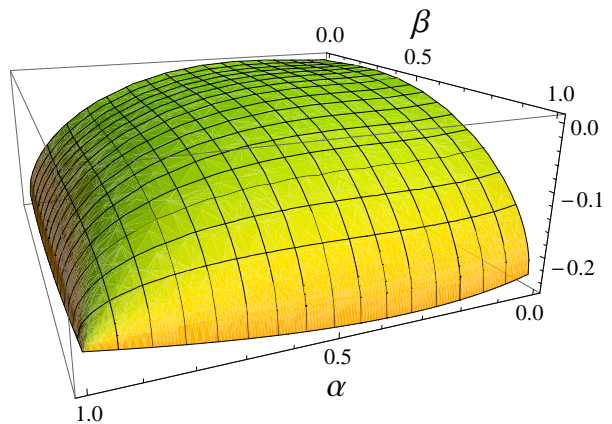}
\caption{The difference $G(\rho_\alpha,\sigma_\beta)-\sqrt{F(\rho_\alpha,\sigma_\beta)}$ 
calculated for states (\ref{eqn:ex-states}) and (\ref{eqn:ex-states-dim4}) as a 
function of parameters $\alpha$ and $\beta$.}
\label{fig:low-bound-diff-dim4}
\end{figure}	

In Fig.~\ref{fig:low-bound-diff} we consider the family $\rho_{\alpha}$ and calculate 
the difference $G(\rho_\alpha,\1/N)-\sqrt{F(\rho_\alpha,\1/N)}$.
One can see that for small dimensions and states close to the
pure state $\ketbra{\psi}{\psi}$ the superfidelity gives a much better approximation
for the trace distance than the fidelity. For larger dimensions this is not the 
case, but nevertheless we can observe that still the superfidelity provides 
a better bound. This advantage is lost for states close to the maximally mixed state.

A similar situation can be observed in Fig.~\ref{fig:low-bound-diff-dim4} where 
the difference between $G(\rho_\alpha,\sigma_\beta)$ and $\sqrt{F(\rho_\alpha,\sigma_\beta)}$ 
($\rho_\alpha,\sigma_\beta\in\Omega_8$) is presented.
For this particular family the bound (\ref{eqn:Main-ineq}) is better than 
(\ref{eqn:fid-dtr-ineq}), but the difference vanishes for states close to the
maximally mixed state.

\begin{figure}[ht!]
\includegraphics[scale=0.7]{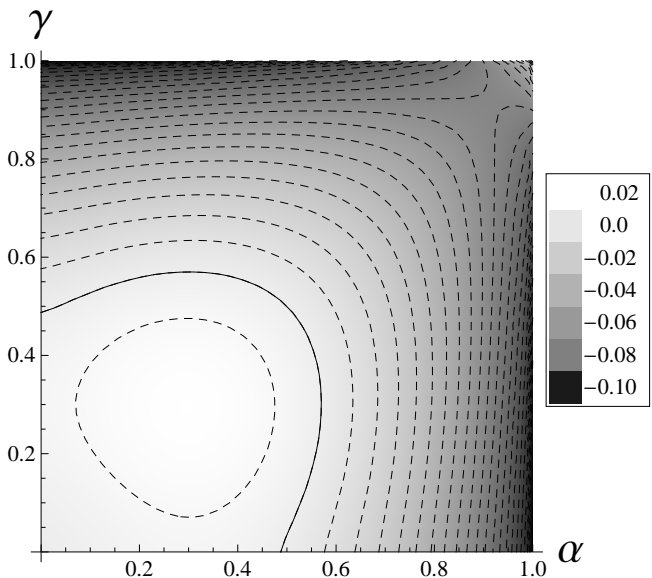}
\caption{The difference $G(\rho_\alpha,\tau_\gamma)-\sqrt{F(\rho_\alpha,\tau_\gamma)}$ 
calculated for states (\ref{eqn:ex-states}) and (\ref{eqn:ex-states-k3}) 
($\rho_\alpha,\tau_\gamma\in\Omega_8$) as the function of parameters 
$\alpha$ and $\gamma$. Solid line is a border that separates 
two regions. For the parameters in lighter one  the inequality (\ref{eqn:fid-dtr-ineq}) 
provides better bound, while in darker region the inequality (\ref{eqn:Main-ineq})
is better.}
\label{fig:diff-levels}
\end{figure}

Unfortunately, the bound (\ref{eqn:Main-ineq}) is not always tighter when compared 
with (\ref{eqn:fid-dtr-ineq}). Figure~\ref{fig:diff-levels} shows the difference 
between $G(\rho_\alpha,\tau_\gamma)$ and $\sqrt{F(\rho_\alpha,\tau_\gamma)}$ 
($\rho_\alpha,\tau_\gamma\in\Omega_8$).
As one can see for this family of states there are regions for which the bound 
(\ref{eqn:Main-ineq}) is better than (\ref{eqn:fid-dtr-ineq}) 
but in this case for highly mixed states 
(\ref{eqn:fid-dtr-ineq}) is better than (\ref{eqn:Main-ineq}).

%%%%%%%%%%%%%%%%%%%%%%%%%%%%%%%%%%%%%%%%%%%%%%%%%%%%%%%%%%%%%%%%%%%%%%%%%%%%%%%%
\section{Final remarks}
%%%%%%%%%%%%%%%%%%%%%%%%%%%%%%%%%%%%%%%%%%%%%%%%%%%%%%%%%%%%%%%%%%%%%%%%%%%%%%%%
We know that the probability of error for distinguishing two density matrices 
$\rho_1,\rho_2\in\Omega_N$ is expressed by the trace distance as \cite{BZ06,ziman}
\begin{equation}
P_E(\rho_1,\rho_2)=\frac{1}{2}[1-D_{\tr}(\rho_1,\rho_2)].
\end{equation}
Using the inequality (\ref{eqn:Main-ineq}) we can write  
\begin{equation}
\frac{1}{2}G(\rho_1,\rho_2)\geq P_E(\rho_1,\rho_2).
\end{equation}

We have shown the relation between the superfidelity and trace distance, which
is analogous to the relation with trace distance and fidelity. This shows
that superfidelity can be used to conclude about the distinguishablity of
states. 

Experimental scheme proposed in \cite{subsuper} can be used to estimate 
superfidelity. Consequently it is possible to estimate experimentally provided 
lower bound for trace distance.

Our bound provides the relation between the trace distance and the overlap of 
two operators supplementary to inequality from \cite[Th.1]{QCB}. As such it 
provides neat mathematical tool which can be used in quantum information theory.

\begin{acknowledgments}
Authors would like to thank K.~\.Zyczkowski, R.~Winiarczyk and P.~Gawron for many 
interesting, stimulating and inspiring discussions.
We acknowledge the financial support by the Polish Ministry of Science and 
Higher Education under the grant number N519 012 31/1957.
\end{acknowledgments}

%%%%%%%%%%%%%%%%%%%%%%%%%%%%%%%%%%%%%%%%%%%%%%%%%%%%%%%%%%%%%%%%%%%%%%%%%%%%%%%%

\end{document}